\documentclass[preprint,titlepage,nofootinbib,superscriptaddress,preprintnumbers,amsmath,amssymb]{revtex4}
\pdfoutput=1

\usepackage[english]{babel}

\usepackage{bm}
\usepackage{hyperref}

\usepackage{graphicx}
\usepackage{float}

\def\iotabar{\lower3pt\hbox{$\mathchar'26$}\mkern-7mu\iota}
\newcommand {\aplt} {\ {\raise-.5ex\hbox{$\buildrel<\over\sim$}}\ }
\newcommand{\dd}{\mbox{d}}

\newcommand{\VZF}{V_{\rm{ZF}}}

\begin{document}

\begin{center}

\end{center}

\title{\bf{Zonal flows and long-distance correlations during the
formation of the edge shear layer in the TJ-II stellarator}}

\author{I. Calvo\footnote{Corresponding author. E-mail: {\tt
ivan.calvo@ciemat.es}}}
\affiliation{Laboratorio Nacional de Fusi\'on,
Asociaci\'on EURATOM-CIEMAT, 28040 Madrid, Spain}

\author{B. A. Carreras}
\affiliation{Universidad Carlos III, 28911 Legan\'es, Madrid, Spain}

\author{L. Garcia}
\affiliation{Universidad Carlos III, 28911 Legan\'es, Madrid, Spain}

\author{M.~A. Pedrosa}
\affiliation{Laboratorio Nacional de Fusi\'on, Asociaci\'on
EURATOM-CIEMAT, 28040 Madrid, Spain}

\author{C. Hidalgo}
\affiliation{Laboratorio Nacional de Fusi\'on, Asociaci\'on
EURATOM-CIEMAT, 28040 Madrid, Spain}


\begin{abstract}
A theoretical interpretation is given for the observed long-distance
correlations in potential fluctuations in TJ-II. The value of the
correlation increases above the critical point of the transition for
the emergence of the plasma edge shear flow layer. Mean (i.e. surface
averaged, zero-frequency) sheared flows cannot account for the
experimental results. A model consisting of four envelope equations
for the fluctuation level, the mean flow shear, the zonal flow
amplitude shear, and the averaged pressure gradient is proposed. It is
shown that the presence of zonal flows is essential to reproduce the
main features of the experimental observations.
\end{abstract}

\maketitle

\section{Introduction}

Transport barrier formation is mostly caused by the emergence of a
radial electric field shear~\cite{Burrell97,Ter00,Burrell06}. This
radial electric field may be induced by poloidal flows and/or a
gradient in the pressure, apart from the direct particle losses.

A simple model for barrier formation and transition to a high
confinement regime that was solely based on the poloidal flow shear
was proposed in Ref.~\cite{DiaLiaCarTer94}. In this model a mean
sheared flow is amplified by the Reynolds
stress~\cite{Hasegawa87,CarLynGar91,DiaKim} and turbulence is
suppressed by shearing~\cite{BigDiaTer}. The combination between those
two effects allows having two possible types of states. On the one
hand, states with vanishing flow shear and high turbulence level (low
confinement). On the other hand, above a critical threshold, states
with non-zero flow shear and reduced turbulence fluctuations (improved
confinement). The transition between these two types of states is a
continuous bifurcation.

In Ref.~\cite{CarNewDiaLia94} the model was extended by incorporating
the pressure gradient component of the radial electric field. This
extended model shows the existence of two critical points, the first
being the same as in the previous model. The second transition,
happening at higher density and temperature, is a discontinuous
transition to a zero fluctuation state where the radial electric field
is only due to the pressure gradient. Experiments have
shown~\cite{Doyle93} that the L to H transition~\cite{WagBecBehetal}
leads to a high confinement state with the radial electric field shear
dominated by the pressure gradient. That is why the second transition
in this model has been associated to the L to H transition.

Recently and in experiments carried out in the TJ-II stellarator, the
first transition (linked to the generation of the poloidal flow) has
been identified~\cite{PedHidCal05,CarGarPedHid06,Pedrosa07,Garcia09}
with the emergence of the plasma edge shear flow
layer~\cite{Ritzetal,Wooton90}.

New experimental results~\cite{Pedrosa08} report the existence of
long-range potential correlations in the toroidal direction. Two
probes are set toroidally separated and not in the same field line. In
addition, consider the intersection point of the field line going
through the first probe with the plane of the second probe. The
distance between this intersection point and the second probe is
larger than a poloidal correlation length of the high-$k$
turbulence. These correlations are observed during the transition
leading to the formation of the plasma edge shear flow layer in
TJ-II. The observed correlations correspond to (non-zero) frequencies
below 30 kHz and thus they cannot be explained by mean (i.e. surface
averaged, zero-frequency) sheared flows. In the present work we aim to
show that the experimental findings of Ref.~\cite{Pedrosa08} can be
understood in the framework of simple transition models if one
appropriately incorporates the contribution of zonal
flows~\cite{DiaItoItoHahm1}. Here we use the term {\it zonal flow} in
the sense of low frequency fluctuating flows with $k_\varphi=0$ and
small but non-zero $k_\theta$. A transition model including zonal
flows was proposed in Ref.~\cite{Kim03}, which we slightly extend here
in order to interpret the TJ-II results. The structure of the model
equations is based on quasilinear calculations from
pressure-gradient-driven turbulence~\cite{Carreras93}. In this paper,
we use a phenomenological approach, determining the main parameters of
the model from the experimental results. We will see that the model is
able to capture the essential features of the experimental
observations.

In Section II we present the transition model incorporating the effect
of zonal flows. In Section III a comparison with the experimental data
is performed. Conclusions are collected in Section IV.

\section{The transition model}

The model presented in this section is an extension of the one used in
Ref.~\cite{CarGarPedHid06} to discuss the emergence of the plasma edge
shear flow layer. This is a model formulated at a radial point. The
dynamical variables are the fluctuation level envelope ${\cal
E}:=\langle(\tilde n/n_0)^2\rangle^{1/2}$, the mean flow shear ${\cal
V}:=\partial_r\langle V_\theta\rangle$, the zonal flow amplitude
shear, ${\cal V}_{\rm{ZF}}:= \partial_r\langle
V_{\theta\rm{ZF}}\rangle$, and (minus) the normalized average pressure
gradient ${\cal N}:=-a\partial_r\langle p\rangle/\langle
p\rangle(0)$. Here $a$ is the minor radius of the torus,
$\langle\cdot\rangle$ stands for the average over angle coordinates,
and $r=0$ corresponds to the magnetic axis. The equations of the model
are
\begin{subequations}\label{eq:4eqModelold}
\begin{align}
&\frac{\dd {\cal E}}{\dd\tau}  = \gamma_0 {\cal N}^{2/3}{\cal E} -
\alpha_1{\cal N}^{-1/2}{\cal E}^2 - \alpha_2{\cal E}{\cal
N}^{-1/3}({\cal V}^2+{\cal V}_{\rm{ZF}}^2),
\label{eq:Eequationold}\\[5pt]
&\frac{\dd  {\cal V}}{\dd\tau} = \bar a_1{\cal N}^{-4/3}{\cal E}^2 {\cal V} +
\bar a_2{\cal N}^{-2/3}{\cal V}_{\rm{ZF}}^2{\cal V} - \bar b{\cal V},
\label{eq:Vequationold}\\[5pt]
&\frac{\dd {\cal V}_{\rm{ZF}}}{\dd\tau}  = \frac{\bar
a_1}{1+\frac{\alpha_2}{\gamma_0}{\cal N}^{-1}{\cal V}^2}{\cal N}^{-4/3}{\cal
E}^2{\cal V}_{\rm{ZF}} + \bar a_3{\cal N}^{-4/3}{\cal E}^2 {\cal V} -
\bar b{\cal V}_{\rm{ZF}},
\label{eq:VZFequationold}\\[5pt]
&\frac{\dd {\cal N}}{\dd\tau} = \bar\Gamma - \bar D{\cal E}{\cal
N}\label{eq:Nequationold}.
\end{align}
\end{subequations}
The structure of these equations is based on a quasilinear
approximation of resistive pressure-gradient driven turbulence (the
resistive interchange mode, due to bad magnetic field line curvature,
is assumed to be the basic instability at the edge of TJ-II). The
linear eigenfunctions and the dependence of the linear growth rates on
$\cal N$ were computed in~\cite{Carreras93}. In particular, we have
used a fluid approach to calculate the poloidal velocity shear and the
sheared radial electric field. The reason is that the TJ-II plasma
edge, $r/a>0.8$, is in the collisional regime. We would like to point
out that for the range of powers and densities in TJ-II considered
here, neoclassical theory is only applicable to the inner region,
$r/a<0.25$. In addition, it has been shown~\cite{Castejon02} that in
this regime the ambipolar radial electric field in TJ-II is small and
shearless for $r>5$ cm and that the electron root~\cite{Hastings85} is
the only accessible root. Therefore, in the edge region the fluid
formulation seems to be adequate for the studies to be carried out in
this paper.

The first term on the right-hand side of Eq.~(\ref{eq:Eequationold})
corresponds to the linear instability generation of turbulence, the
second to the non-linear saturation of the instability, and the two
last terms to the supression of turbulence by sheared mean and zonal
flows. The first and second terms on the right-hand side of
Eq.~(\ref{eq:Vequationold}) represent the generation of mean sheared
flow by Reynolds stress, and the third one is the collisional damping
term (analogous to the last term on the right-hand side of
Eq.~(\ref{eq:VZFequationold})). The two first terms on the right-hand
side of Eq.~(\ref{eq:VZFequationold}) give the generation of zonal
flow by Reynolds stress; in the first term the factor $(1+\alpha_2
{\cal N}^{-1}{\cal V}^2 /\gamma_0)^{-1}$ corresponds to the effect of
zonal flow supression by a mean sheared flow. Finally, in
Eq.~(\ref{eq:Nequationold}), $\bar D{\cal E}$ is the anomalous
particle diffusivity, and $\bar\Gamma$ the normalized incremental
particle flux, the control parameter of the model. Diamagnetic effects
in the momentum balance equation have been neglected because we
consider small values of $\bar\Gamma$.

In terms of dimensionless variables,
\begin{equation}
t = \gamma_0\tau, \ \ E = \frac{\alpha_1}{\gamma_0}{\cal E}, \ \
V = \sqrt{\frac{\alpha_2}{\gamma_0}}{\cal V}, \ \ \VZF =
\sqrt{\frac{\alpha_2}{\gamma_0}}{\cal V}_{\rm{ZF}}, \ \ N = {\cal N},
\end{equation}
the equations read:
\begin{subequations}\label{eq:4eqModel}
\begin{align}
&\frac{\dd E}{\dd t} = N^{2/3}E - N^{-1/2}E^2 - N^{-1/3}E(V^2+\VZF^2),
\label{eq:Eequation}\\[5pt]
&\frac{\dd V}{\dd t} = a_1N^{-4/3}E^2 V + a_2N^{-2/3}\VZF^2V-bV,
\label{eq:Vequation}\\[5pt]
&\frac{\dd \VZF}{\dd t} =
\frac{a_1}{1+N^{-1}V^2}N^{-4/3}E^2\VZF+a_3N^{-4/3}E^2 V - b\VZF,
\label{eq:VZFequation}\\[5pt]
&\frac{\dd N}{\dd t} = \Gamma -
DEN\label{eq:Nequation}.
\end{align}
\end{subequations}
where $a_1 = \gamma_0\bar a_1/\alpha_1^2$, $a_2 = \bar a_2/\alpha_2$,
$a_3 = \gamma_0\bar a_3/\alpha_1^2$, $b = \bar b/\gamma_0$, $\Gamma =
\bar\Gamma/\gamma_0$, and $D = \bar D/\alpha_1$.

The form of the equation for the time evolution of zonal flows,
Eq.~(\ref{eq:VZFequation}), coincides with the one proposed in
Ref.~\cite{Kim03}, except for the term proportional to $a_3$, which is
absent in the latter reference. It is worth commenting on the physical
origin of that term. In the framework of the paradigm of shear flow
generation by turbulence the Reynolds stress gives a non-zero
contribution when the turbulent eddies are distorted by the presence
of global shear flows. If only the mean flow is present the Reynolds
stress gives the mean shear flow amplification term that we have
discussed in the past~\cite{DiaLiaCarTer94}. When, in addition, zonal
flows exist, the Reynolds stress gives two main contributions to the
zonal flow equation. One comes from the coupling of $m$ and $-m+q$
components of the eigenfunctions distorted by the zonal flow (the first
term on the rhs of Eq.~(\ref{eq:VZFequation})). The other comes from a
similar coupling but with the distortion induced by the mean flow
(the second term on the rhs of Eq.~(\ref{eq:VZFequation})). Here $m$
is large and corresponds to the turbulent component of the flow,
whereas $q$ is the wave-number of the zonal flow. A detailed
calculation of those terms will be provided elsewhere.

As will be shown below, there is a qualitative difference
between $a_3=0$ and $a_3\neq 0$. If $a_3=0$ the model exhibits a
continuous transition between the state with $V=0$ and the state with
$V\neq 0$. In addition, the stable fixed points are such that
$\VZF=0$. However, if $a_3\neq 0$ the transition is discontinuous and
the stable, improved confinement state has both $V$ and $\VZF$
non-vanishing. Since the toroidal correlations will be associated to
the existence of stationary zonal flows, it seems that small but
non-zero $a_3$ is required. Obviously, for very small $a_3$ it is not
possible to directly (that is, according to the continuity of the
order parameter) distinguish between a continuous and a discontinuous
transition.

\subsection{The toroidal correlation}

In this subsection we will try to express the correlation of the
potential fluctuations at two toroidal positions separated by a
toroidal angle $\delta$ in terms of the variables of our model. The
formula defining the correlation is
\begin{equation}\label{crosscorr}
\mu = \frac{\Big\langle\Big(\Phi(r,\theta,\varphi,t)-
\langle\Phi(r,\theta,\varphi,t)\rangle\Big)
\Big(\Phi(r,\theta,\varphi+\delta,t)-
\langle\Phi(r,\theta,\varphi+\delta,t)\rangle\Big)\Big\rangle}{\sqrt{\Big\langle\Big(\Phi(r,\theta,\varphi,t)-
\langle\Phi(r,\theta,\varphi,t)\rangle\Big)^2\Big\rangle
\Big\langle\Big(\Phi(r,\theta,\varphi+\delta,t)-
\langle\Phi(r,\theta,\varphi+\delta,t)\rangle\Big)^2\Big\rangle}}.
\end{equation}

Assume that a separation of time scales exists, so that one can write
\begin{equation}
\Phi(r,\theta,\varphi,t)- \langle\Phi(r,\theta,\varphi,t)\rangle =
\Phi_{\rm{ZF}}(r,\theta,t) + \tilde\Phi(r,\theta,\varphi,t),
\end{equation}
where $\Phi_{\rm{ZF}}(r,\theta,t)$ is related to the zonal flow and
$\tilde\Phi(r,\theta,\varphi,t)$ to high frequency turbulent
fluctuations. The high frequency fluctuations have short correlation
length in the toroidal direction, except when the positions are
aligned with the field lines. Here we assume that this is never the
case. The zonal flows are characterized by a low frequency and not
having a dependence on the toroidal angle. Now, take a field line on
the magnetic surface labeled by $r$ passing through $(\theta,\varphi)$
and $(\theta_\delta,\varphi +\delta)$. We are assuming that $\delta$
is such that $|r(\theta_\delta - \theta)|\gg l_\theta$, where
$l_\theta$ is the poloidal correlation length of the high-$k$
turbulence.

For us, $\langle\cdot\rangle$ denotes an average over $\theta$ and
$\varphi$. Eq.~(\ref{crosscorr}) takes the form:
\begin{equation}\label{eq:crosscorr2}
\mu = \frac{\langle\Phi_{\rm{ZF}}(r,\theta,t)^2\rangle}
{\langle\Phi_{\rm{ZF}}(r,\theta,t)^2\rangle +
\langle\tilde\Phi(r,\theta,\varphi,t)^2\rangle} = \frac{1} {1 + \frac{
\langle\tilde\Phi(r,\theta,\varphi,t)^2\rangle}
{\langle\Phi_{\rm{ZF}}(r,\theta,t)^2\rangle}},
\end{equation}
where we used that the toroidal correlation of turbulent fluctuations,
\begin{equation}
\langle\tilde\Phi(r,\theta,\varphi,t)
\tilde\Phi(r,\theta,\varphi+\delta,t)\rangle,
\end{equation}
is negligible for $\delta$ large enough.

The scenario suggested by the above considerations in order to
interpret the experimental results of Ref.~\cite{Pedrosa08} is
clear. We expect that in a ramping experiment crossing the critical
point, $\langle\Phi_{\rm{ZF}}^2\rangle /
\langle\tilde\Phi^2\rangle $ be zero below the
critical point and non-zero above it. From Eq.~(\ref{eq:crosscorr2})
we deduce that this makes the correlation function, $\mu$, grow during
the transition.

Let us finally write (\ref{eq:crosscorr2}) in terms of the variables
of the present model. Since the model equations can be derived from
quasilinear calculations of a pressure-gradient-driven turbulence
model (which in particular is a fluid model), we assume that the
density perturbation is the result of the convection of the
equilibrium density by the flow ${\bf
\tilde V}=-{\nabla\tilde\Phi}\times{\bf B}/B^2$. That is,
\begin{equation}
{\tilde\Phi}_k \approx \frac{\gamma_k{\tilde n}_k}{k_\theta dn_0/dx}. 
\end{equation}
Denote by $\{\cdot\}$ the spectrum average. Using that
$\langle\tilde\Phi^2\rangle\propto \{|{\tilde\Phi}_k|^2\}$,
and $\gamma_k\propto N^{2/3}$ (see Ref.~\cite{Carreras93}), we infer that
\begin{equation}
\langle\tilde\Phi^2\rangle \propto N^{-2/3}E^2.
\end{equation}
Also, $\VZF\propto \Phi_{\rm{ZF}}$. Therefore,
\begin{equation}
\frac{\langle\tilde\Phi^2\rangle}{\langle\Phi_{\rm{ZF}}^2\rangle}
= \lambda\frac{E^2}{N^{2/3}\VZF^2}, \quad \lambda>0.
\end{equation}
Hence,
\begin{equation}\label{eq:formulamu}
\mu = \left(1 + \lambda\frac{E^2}{N^{2/3}\VZF^2}\right)^{-1}.
\end{equation}

This is the formula we were looking for. It gives the toroidal
correlation of the electrostatic potential in terms of the variables
of our model. In particular, it shows in a manifest way that the zonal
flow is responsible for the appearance of toroidal correlations.

\subsection{Fixed points}

The fixed points of Eqs.~(\ref{eq:4eqModel}) are the solutions of
\begin{subequations}\label{eq:4eqModelFixed}
\begin{align}
&N^{2/3}E - N^{-1/2}E^2 -  N^{-1/3}E(V^2+\VZF^2)=0, \label{eq:EeqFixed}\\
&(a_1N^{-4/3}E^2 + a_2N^{-2/3}\VZF^2-b)V=0, \label{eq:VeqFixed}\\
&\left(\frac{a_1}{1+N^{-1}V^2}N^{-4/3}E^2 - b\right)\VZF + a_3N^{-4/3}E^2 V= 0,
\label{eq:VZFeqFixed}\\
&DEN=\Gamma\label{eq:NeqFixed}.
\end{align}
\end{subequations}

A fixed point corresponding to a low confinement regime always
exists:
\begin{itemize}
\item[(i)] $V_0 = \VZF{}_0 = 0,\ E_0 = (\Gamma/D)^{7/13}, \ N_0
=(\Gamma/D)^{6/13}$.
\end{itemize}
Define the critical flux, $\Gamma_c:=D(a_1/b)^{-13/6}$. The fixed
point (i) is stable if $\Gamma<\Gamma_c$ and unstable if
$\Gamma>\Gamma_c$.

It is easy to see that when $\Gamma>\Gamma_c$, there is another fixed
point:
\begin{itemize}
\item[(ii)] $V_0 = 0$, $\VZF{}_0^2=N-N^{-1/6}E =
(a_1/b)^{3/10}(\Gamma/D)^{3/5}- (a_1/b)^{-7/20}(\Gamma/D)^{3/10}$, $E_0
= (b/a_1)^{3/10}(\Gamma/D)^{2/5}$, $N_0 = (a_1/b)^{3/10}(\Gamma/D)^{3/5}$,
\end{itemize}
which is always unstable.

\begin{figure}[H]
\centering
\includegraphics[angle=0,width=10cm]{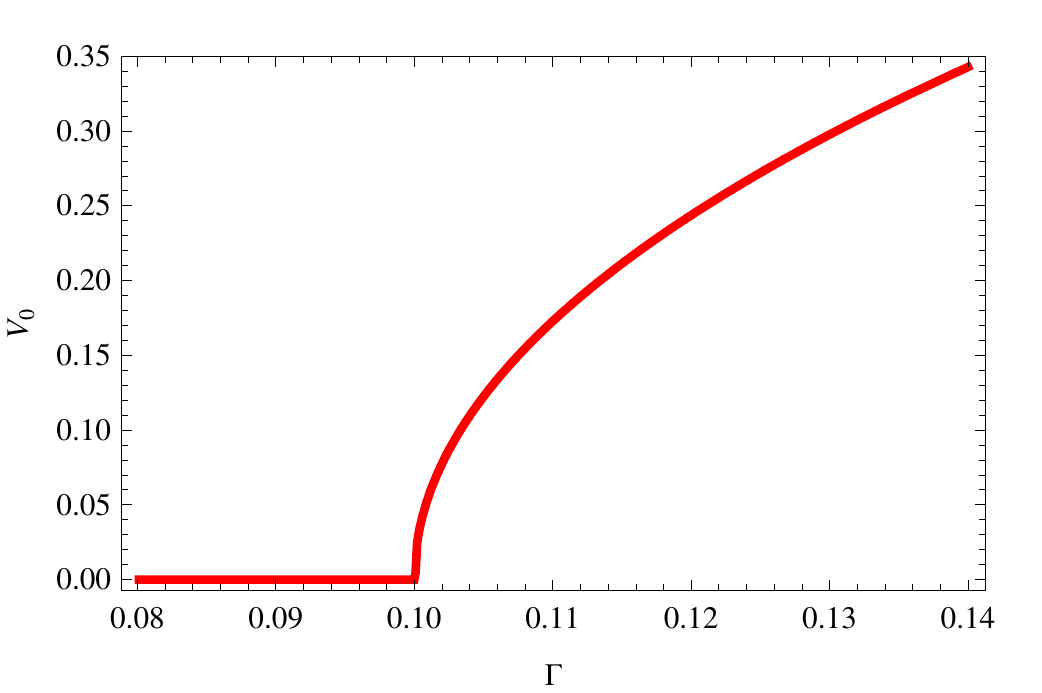}
\caption{Stable stationary values of $V$. The values of the parameters
are $a_1=b=D=0.1$, $a_2=0.5$, $a_3=0$.}
\label{FIG:Fig_Bifurc2order}
\end{figure}

The discussion of the fixed point with $V\neq 0$ is more difficult and
depends on the value of $a_3$. If $a_3=0$ and $\Gamma>\Gamma_c$ there
is an additional fixed point
\begin{itemize}
\item[(iii)] $\VZF{}_0 = 0$, $V_0^2=N-N^{-1/6}E =
(a_1/b)^{3/10}(\Gamma/D)^{3/5}- (a_1/b)^{-7/20}(\Gamma/D)^{3/10}$, $E_0
= (b/a_1)^{3/10}(\Gamma/D)^{2/5}$, $N_0 = (a_1/b)^{3/10}(\Gamma/D)^{3/5}$,
\end{itemize}
which is stable\footnote{At least for moderate values of
$\Gamma$.}. Note that the transition is continuous at
$\Gamma=\Gamma_c$. A plot of the bifurcation is given in
Fig.~\ref{FIG:Fig_Bifurc2order}.

\begin{figure}[H]
\centering
\includegraphics[angle=0,width=10cm]{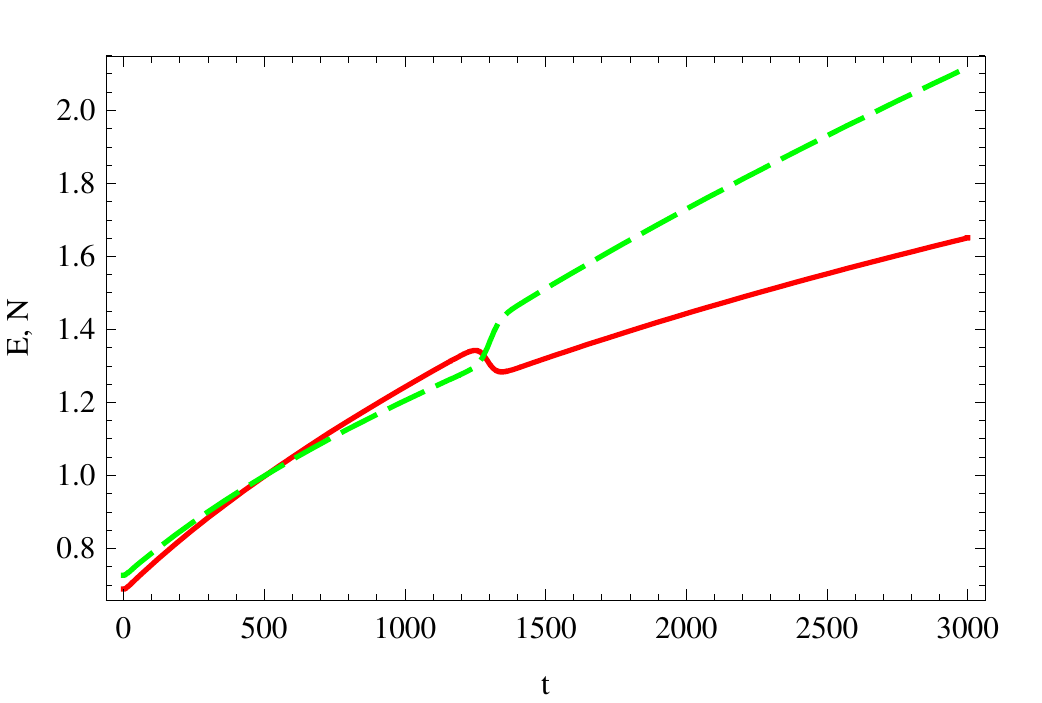}
\includegraphics[angle=0,width=10cm]{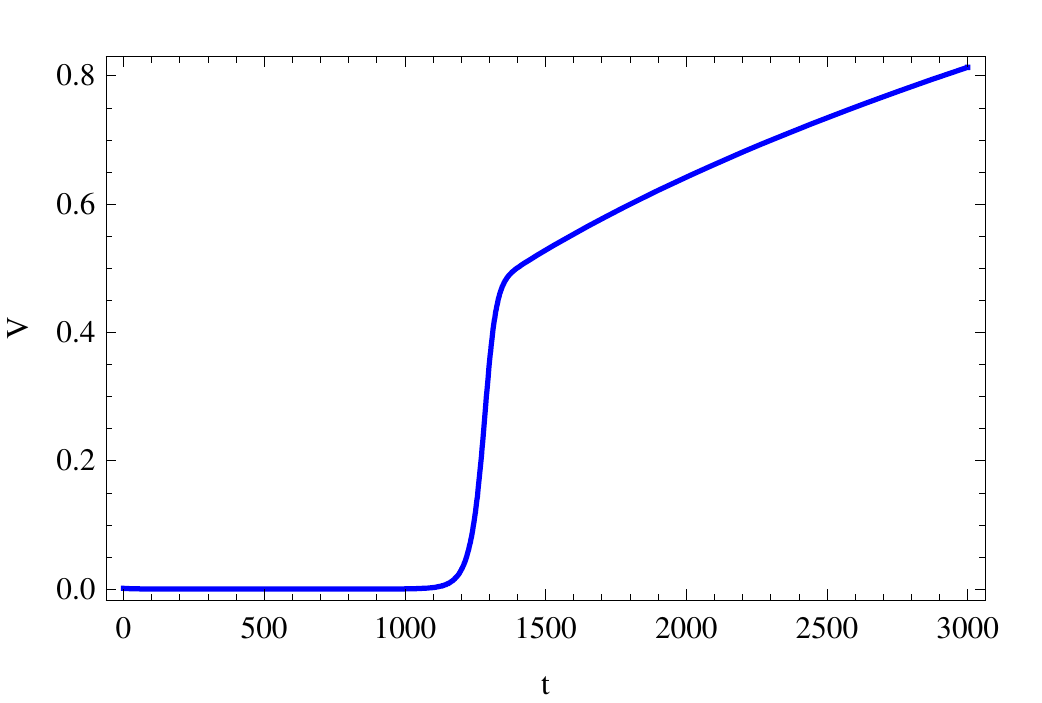}
\includegraphics[angle=0,width=10cm]{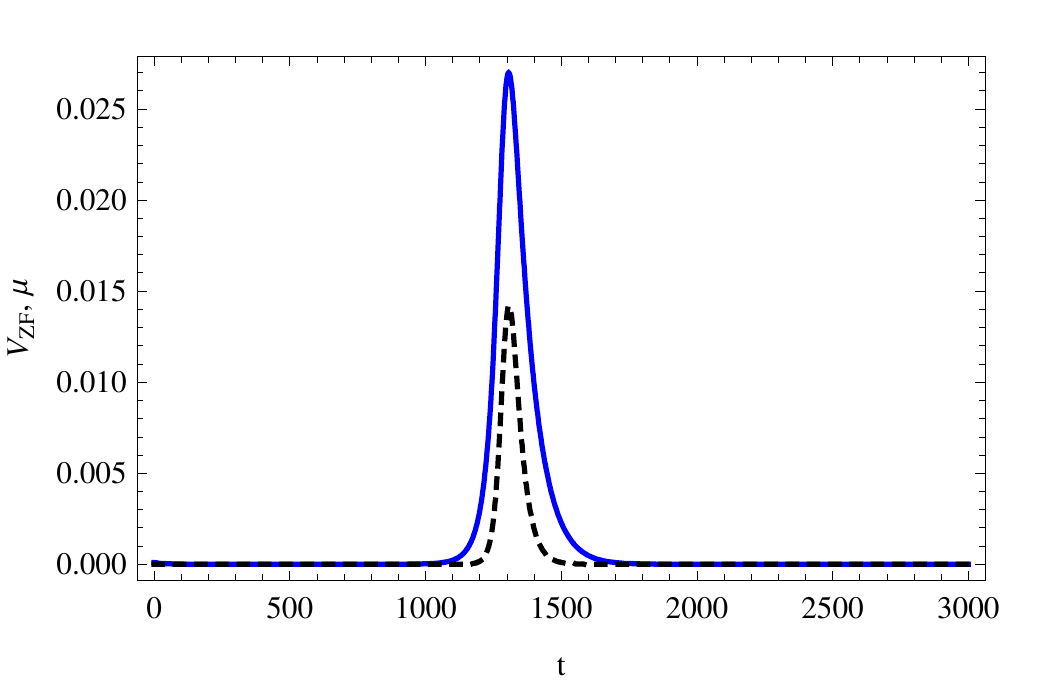}
\caption{Time-evolution of the variables of the model for
$\Gamma(t)=0.05 + 10^{-4}t$, $a_1=b=D=0.1$, $a_2=0.5$, and $a_3=0$.
Top: $E$ (solid) and $N$ (dashed). Middle: $V$. Bottom: $\VZF$ (solid)
and $\mu$ (dashed). The initial conditions are $V(0)=10^{-3}$,
$\VZF(0)=10^{-4}$, $E(0)=(\Gamma(0)/D)^{7/13}$,
$N(0)=(\Gamma(0)/D)^{6/13}$.}
\label{FIG:Fig_TipicRun_a30}
\end{figure}

When $a_3=0$, the dynamics never reaches an equilibrium solution with
$\VZF\neq 0$. Equivalently, $\VZF$ can be non-zero only
transiently. But this is a problem for reproducing the long-range
correlations observed during the transition to improved-confinement
regimes in TJ-II. The results of Ref.~\cite{Pedrosa08} were obtained
in ramping experiments traversing the critical point. In
Fig.~\ref{FIG:Fig_TipicRun_a30} we show the numerical results from our
model when one performs a flux ramp traversing the critical point,
going from a low confinement state to an improved confinement one. When
$V$ starts growing from zero, and during a short time, $\VZF$ follows
it and becomes non-zero. However, at a certain moment the inhibition
of $\VZF$ by $V$ becomes noticeable as $V$ increases and $\VZF$
decreases to zero after the transition. The interval of time in which
$\VZF\neq 0$ coincides with the interval in which $\mu\neq
0$. However, in the experimental data one can see that the correlation
has a non-zero stationary value above the critical point.

\begin{figure}[H]
\centering
\includegraphics[angle=0,width=10cm]{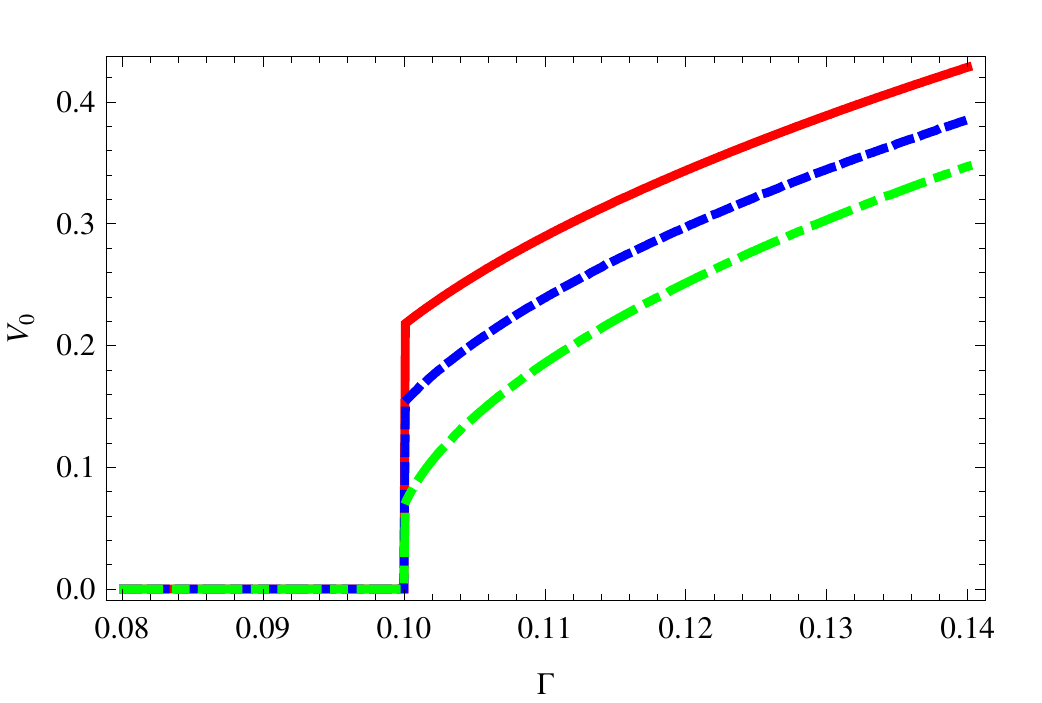}
\caption{Stable stationary values of $V$. The values of the
parameters are $a_1=b=D=0.1$, $a_2=0.5$, and $a_3=0.01$ (solid),
$a_3=0.005$ (dashed), $a_3=0.001$ (long-short dash).}
\label{FIG:Fig_Bifurc1orderV}
\end{figure}

\begin{figure}[H]
\centering
\includegraphics[angle=0,width=10cm]{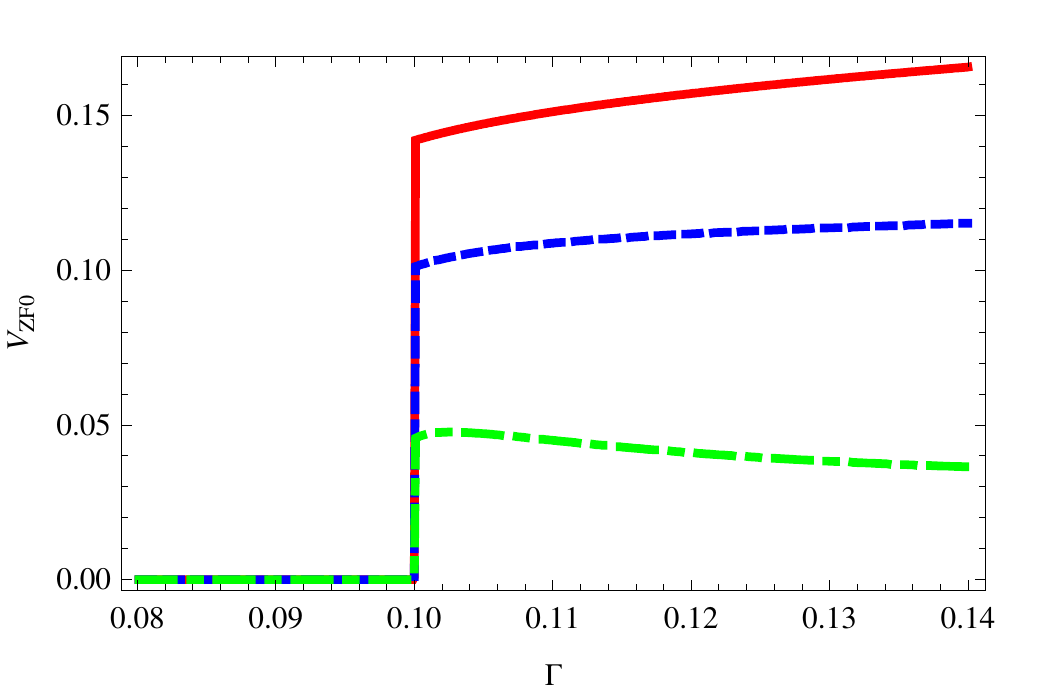}
\caption{Stable stationary values of $\VZF$. The values of the
parameters are $a_1=b=D=0.1$, $a_2=0.5$, and $a_3=0.01$ (solid),
$a_3=0.005$ (dashed), $a_3=0.001$ (long-short dash).}
\label{FIG:Fig_Bifurc1orderVzf}
\end{figure}

\begin{figure}[H]
\centering
\includegraphics[angle=0,width=10cm]{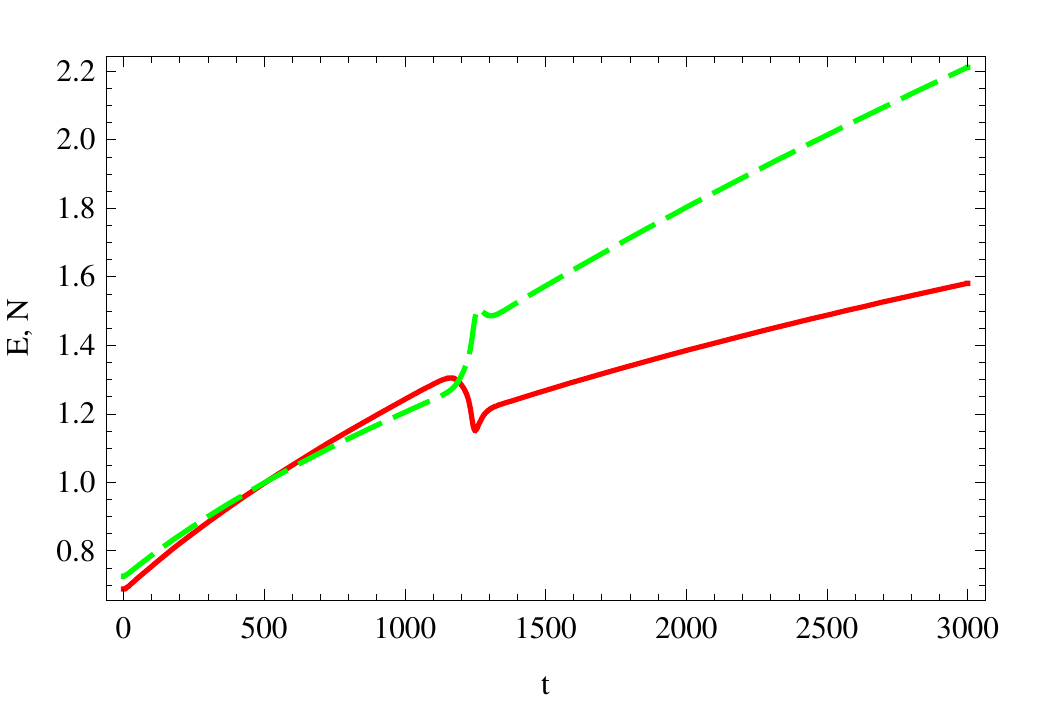}
\includegraphics[angle=0,width=10cm]{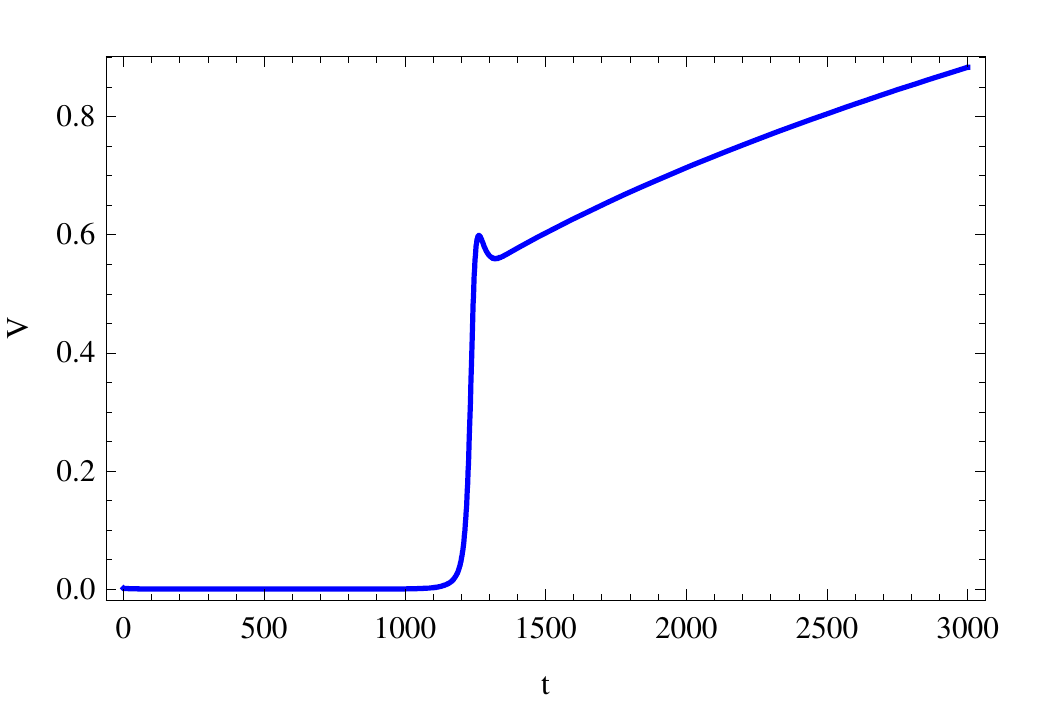}
\includegraphics[angle=0,width=10cm]{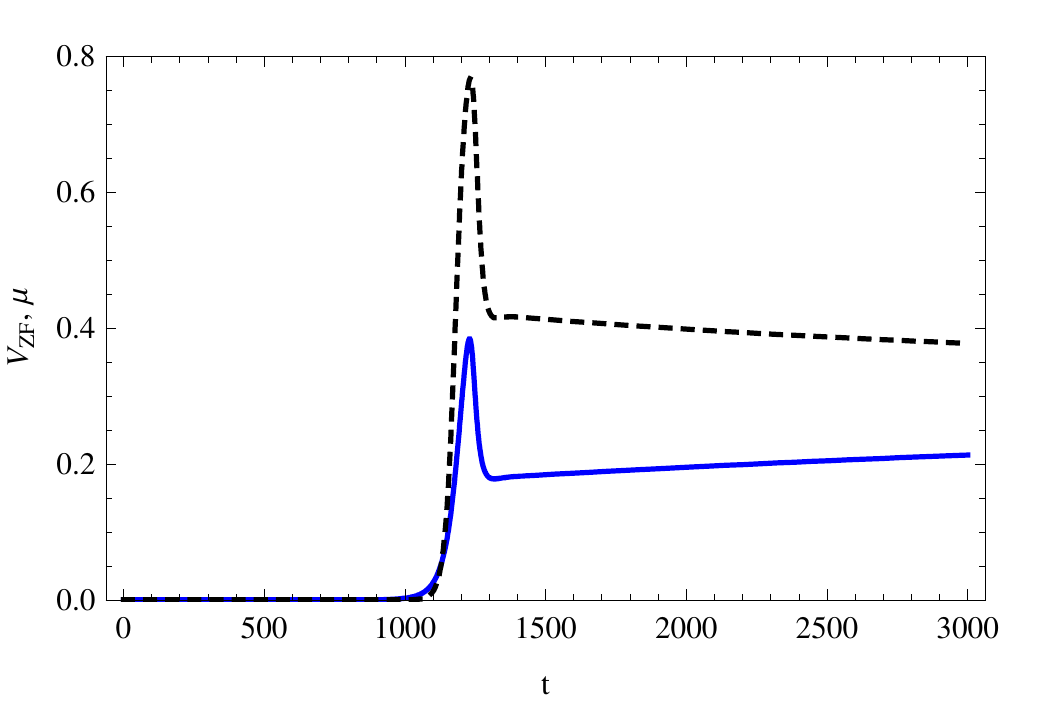}
\caption{Time-evolution of the variables of the model for the same
values of the parameters as in Fig.~\ref{FIG:Fig_TipicRun_a30}, except
that $a_3=0.01$.}
\label{FIG:Fig_TipicRun_a31e-2}
\end{figure}

The situation is very different for $a_3>0$. The fixed points (i) and
(ii) and their linear stability remain unchanged. However, there is no
solution with $\VZF=0$ and $V\neq 0$. The third fixed point,
corresponding to the high confinement regime and which we will call
(iii)$'$, has both $V$ and $\VZF$ non-zero. In
Figs.~\ref{FIG:Fig_Bifurc1orderV} and \ref{FIG:Fig_Bifurc1orderVzf} we
show numerical calculations of the bifurcation for different values of
$a_3$. The most remarkable feature is that for $a_3\neq 0$ the
transition is not continuous anymore, but the stationary values of the
variables jump at $\Gamma_c$. Of course, the magnitude of the jump
decreases when $a_3$ decreases. Regarding the problem of long-range
correlations, it seems essential to have non-zero $a_3$. As shown in
Fig.~\ref{FIG:Fig_TipicRun_a31e-2} this allows to have rampings in
which the correlations reach a non-zero stationary value. Although
there is no simple analytical expression for the fixed point (iii)$'$
for a general value of $\Gamma$, we can give a good approximation for
the supercritical solution at $\Gamma = \Gamma_c$. Define $\epsilon: =
\sqrt{a_3}$ and assume
\begin{subequations}\label{eq:Perturb_a3}
\begin{align}
& E = E_c + \epsilon^2 E_2 + o(\epsilon^3), \quad N = N_c + \epsilon^2
N_2 + o(\epsilon^3),\\
& V = \epsilon V_1 + o(\epsilon^2), \quad \VZF =
\epsilon V_{\rm{ZF}1} + o(\epsilon^2).
\end{align}
\end{subequations}
$E_c = (b/a_1)^{7/6}$ and $N_c = b/a_1$ are the stationary values of
$E$ and $N$ for $a_3=0$. Introducing this expansion in
Eqs.~(\ref{eq:4eqModelFixed}) and solving for the lowest order we get:

\begin{subequations}\label{eq:Perturb_a3Solution}
\begin{align}
& E_2 = -\frac{3a_2}{\sqrt{5}\ a_1}\left(\frac{b}{a_1}\right)^{5/6}
\frac{\sqrt{13\frac{a_2}{a_1}-20\left(\frac{b}{a_1}\right)^{2/3}}}{33a_2-20a_1
\left(\frac{b}{a_1}\right)^{2/3}} \\ & N_2 =
-\left(\frac{b}{a_1}\right)^{-1/6} E_2 \\ & V_1^2 =
\left(-\frac{13}{6}\left(\frac{b}{a_1}\right)^{-1/6} +
\frac{10}{3}\frac{a_1}{a_2}\left(\frac{b}{a_1}\right)^{1/2}\right)E_2\\
& V_{\rm{ZF}1}^2 =
-\frac{10}{3}\frac{a_1}{a_2}\left(\frac{b}{a_1}\right)^{1/2}E_2.
\end{align}
\end{subequations}

At this point, we must comment on an issue. In the experiments, that
are carried out by means of density ramps, it is difficult to
distinguish between a continuous and a discontinuous transition. We
can see that by comparing the averaged flows in
Figs.~\ref{FIG:Fig_TipicRun_a30} and \ref{FIG:Fig_TipicRun_a31e-2}. A
continuous transition may appear very sharp because the velocity shear
grows exponentially in the initial phase. The sharpness depends on the
value of this growth rate and on the noise level from where the
velocity emerges.

\section{Comparison of model and experiment}

Experiments were carried out in the TJ-II stellarator in Electron
Cyclotron Resonance Heated plasmas ($P_{{\rm ECRH}}\leq 400$ kW,
$B_T=1$ T, $\langle R \rangle = 1.5$ m, $\langle a \rangle \leq 0.22$
m, $\iotabar(a)\in [1.5,1.9]$). The plasma density was varied in the
range $[0.35\cdot 10^{19},1\cdot 10^{19}]{\rm m}^{-3}$. Different edge
plasma parameters were simultaneously characterized in two different
toroidal positions approximately $160^\circ$ apart using two similar
multi-Langmuir probes, installed on fast reciprocating drives
(approximately $1$ m/s)~\cite{Pedrosa99}. For details on the probe
arrangement see Ref.~\cite{Pedrosa08}. Probe 1 is located in a top
window entering vertically through one of the `corners' of its
beam-shaped plasma and at $\varphi\approx 35^\circ$ (where $\varphi$
is the toroidal angle in the TJ-II reference system). Probe 2 is
installed in a bottom window at $\varphi\approx 195^\circ$ and enters
into the plasma through a region with a higher density of flux
surfaces (i.e. lower flux expansion) than Probe 1. It is important to
note that the field line passing through one of the probes is
approximately $100^\circ$ poloidally apart when reaching the toroidal
position of the other probe that is more than $5$ m away. Edge radial
profiles of different plasma parameters have been measured
simultaneously at the two separated toroidal locations with very good
agreement. Profiles were obtained in both shot to shot and single shot
scenarios with the two probes and in different plasma configurations.

There are five parameters in the model and the $\lambda$ parameter in
the determination of the correlation; this is apart from the input
function $\Gamma$. The two parameters $b$ and $D$ are directly related
to the dissipation terms, viscosity and transport. We determine them
from the expected values of those terms at the plasma edge. We take
the flow damping rate to be $\nu_{ii} = 1.7\cdot 10^4 \ \rm{s}^{-1}$ and
the particle diffusivity $D_p = 10^5 \ \rm{cm}^2\rm{s}^{-1}$.
Therefore, we used $b=D=0.1$. The $a_1$ parameter is determined from
the criticality condition and the experimental measure of the density
profile at the critical point. The measured density profile at about
the critical density in TJ-II is such that $N_c \approx 1$. Therefore,
we take $a_1=0.1$.

In this model the long-range correlations are controlled essentially
by the ratio $a_3/\lambda$. Observe that using
Eqs.~(\ref{eq:Perturb_a3}) and (\ref{eq:Perturb_a3Solution}) we can
find an approximate value of the correlation at the critical point:
\begin{equation}
\mu\approx
\left(1+\frac{\lambda}{a_3}\frac{N_c^{-2/3}E_c^2}{V_{\rm{ZF}1}^2}\right)^{-1}.
\end{equation}

To have a reasonable level of correlation we need $a_3/\lambda$ about
1/3. With the present data it is not possible to distinguish between
the two parameters. Therefore, just for convenience,we have taken
$a_3=0.01$ and $\lambda=0.03$. Finally, the parameter $a_2$ has not a
very visible impact on the comparison with the data and we have chosen
$a_2=0.5$.

\begin{figure}[H]
\centering
\includegraphics[angle=0,width=10cm]{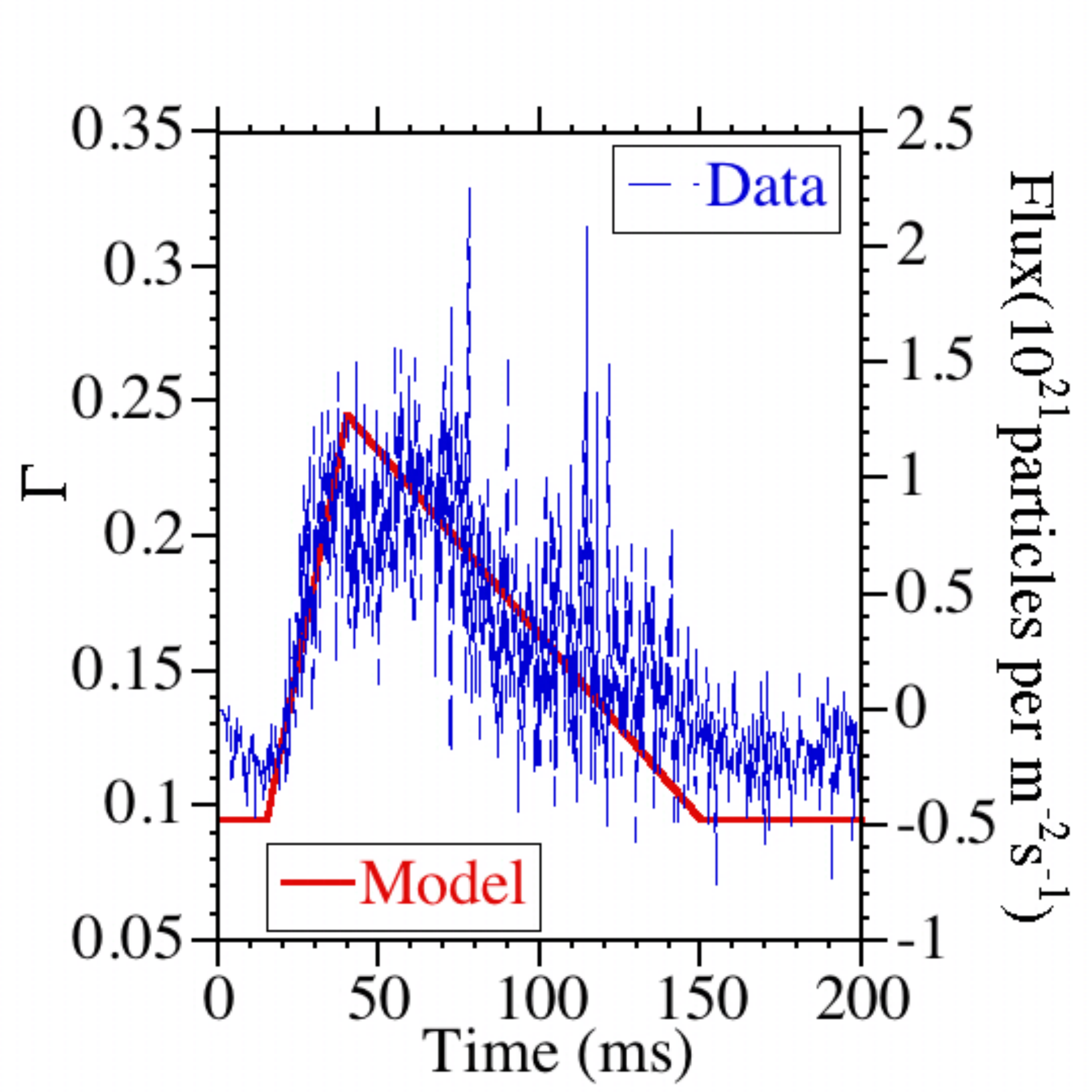}
\caption{Flux function, $\Gamma(t)$, used in modeling the experimental
data.}
\label{FIG:Fig_ExperimFlux}
\end{figure}

\begin{figure}[H]
\centering
\includegraphics[angle=0,width=8cm]{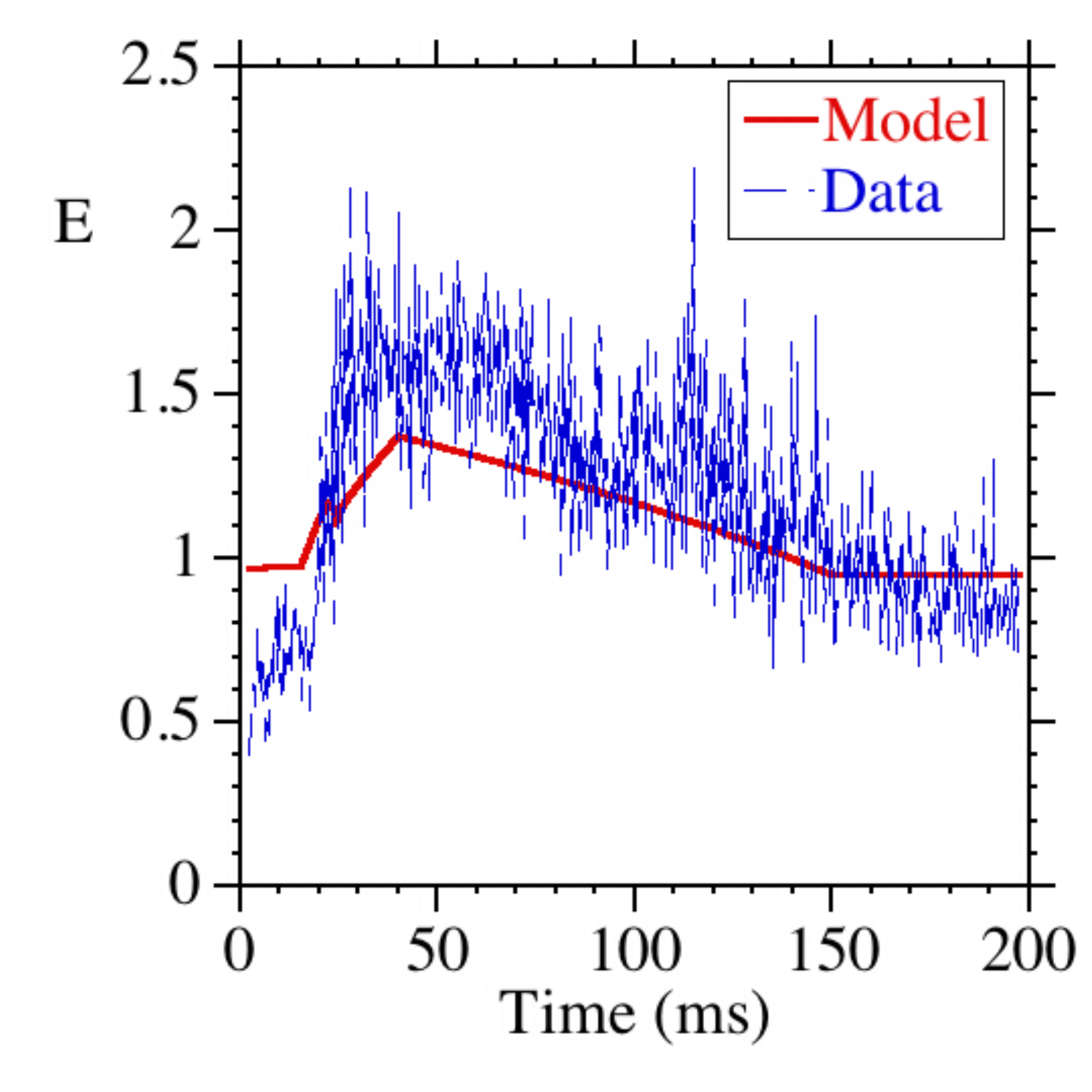}
\includegraphics[angle=0,width=8cm]{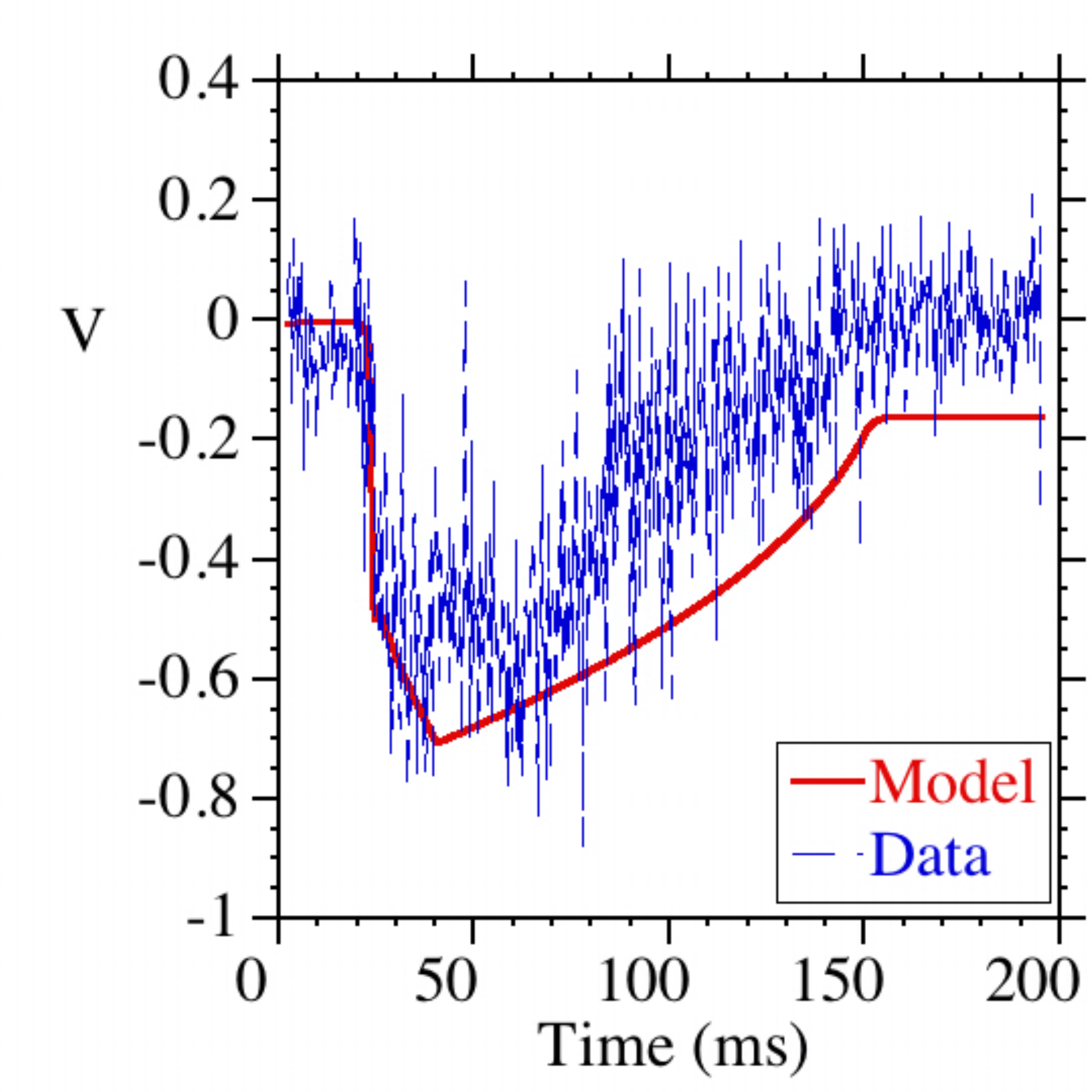}
\includegraphics[angle=0,width=8cm]{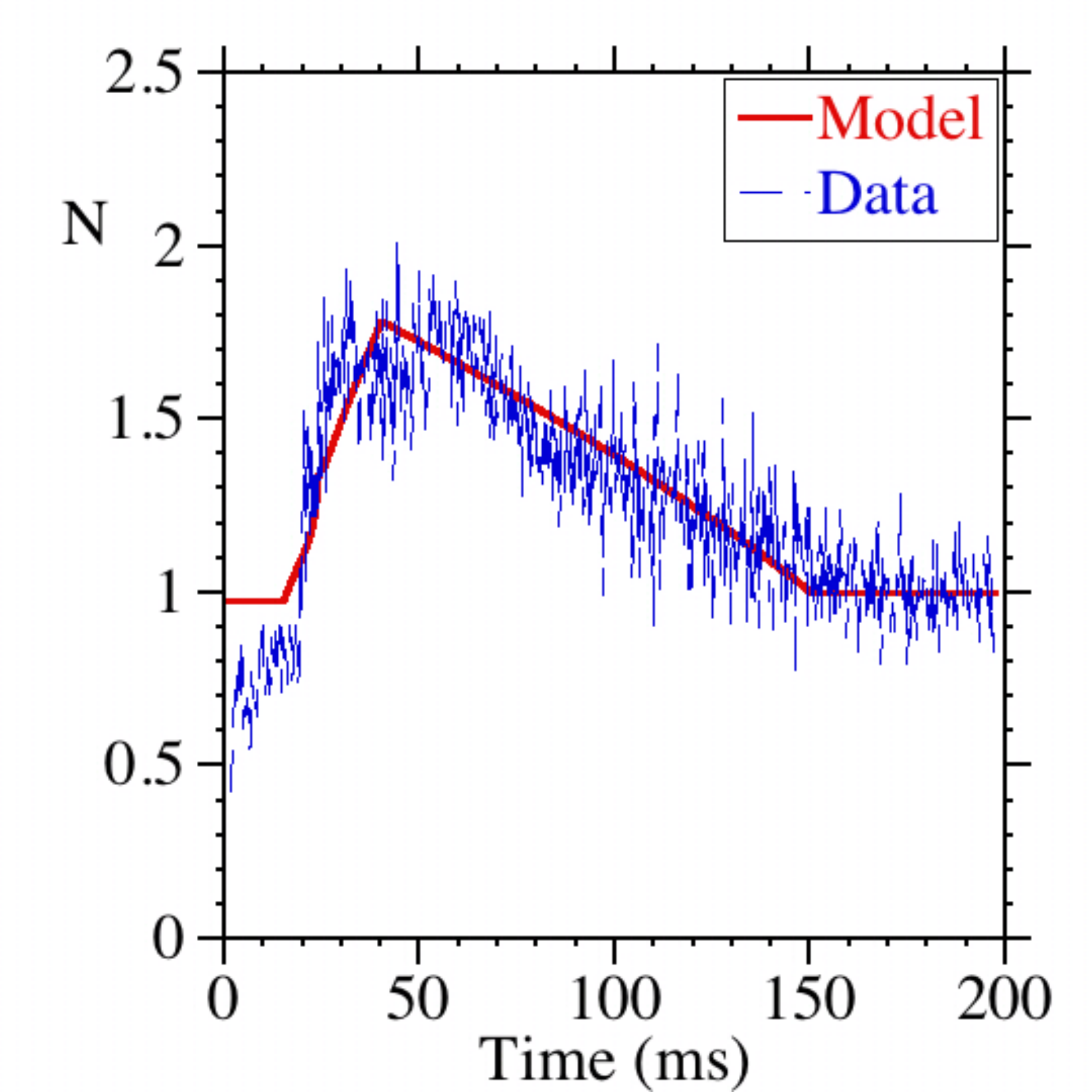}
\includegraphics[angle=0,width=8cm]{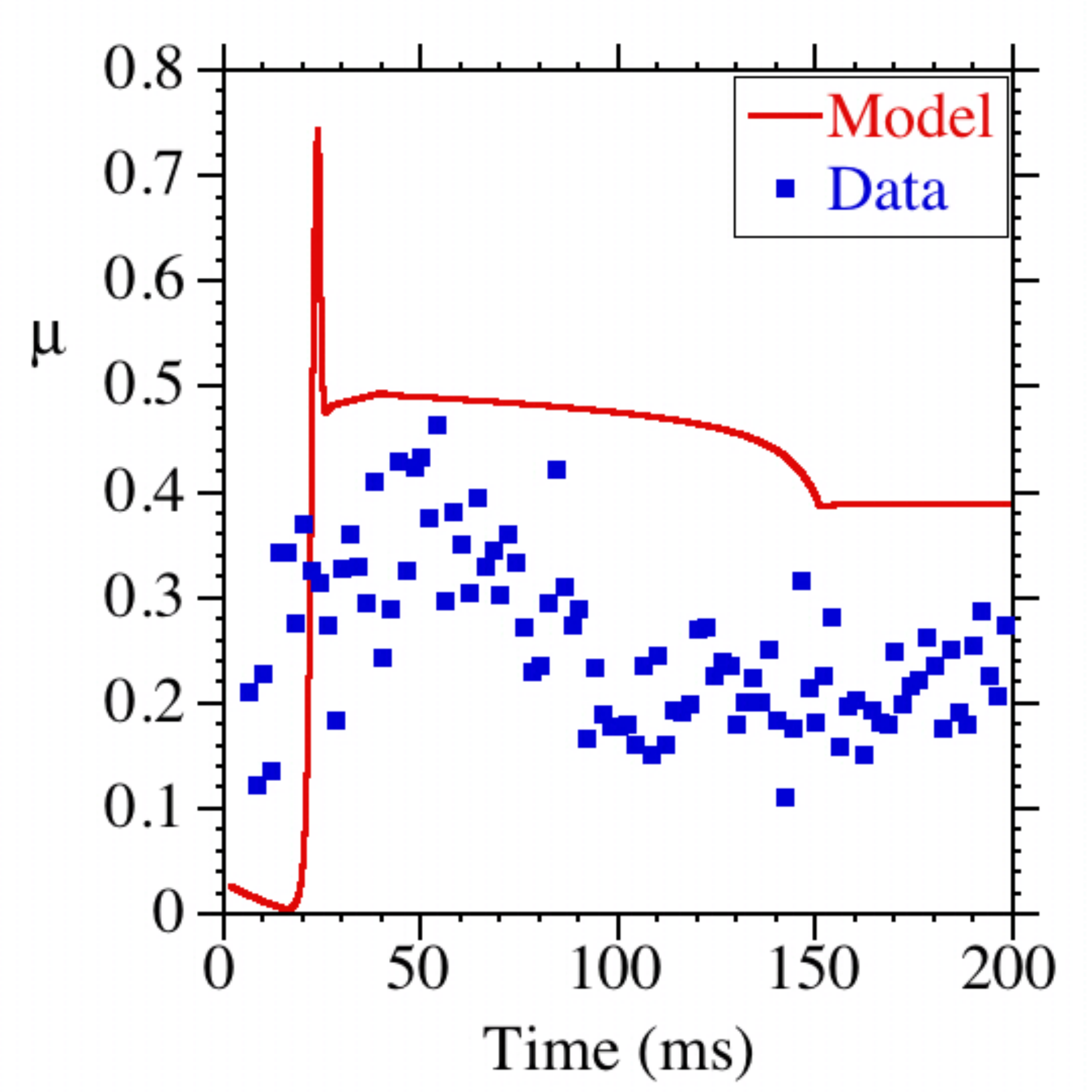}
\caption{Comparison between the model equations (\ref{eq:4eqModel})
and the experimental data for the shot 18229.}
\label{FIG:Fig_CompTheoExp}
\end{figure}

The input function required in modeling each discharge is the flux
function $\Gamma(t)$. Since we are not doing a detailed modeling of
data, but only a description of the main features, we have
parameterized the flux using only linear dependences in time. A
typical example of how the data are described by the model is shown in
Figs.~\ref{FIG:Fig_ExperimFlux} and \ref{FIG:Fig_CompTheoExp},
corresponding to discharge 18229. This is a case with a ramp up and
down where the plasma crosses the critical point twice, once in the
way up and another in the way down.  Similar results have been
obtained for 10 discharges of TJ-II using the same set of values for
the parameters. As one can see in Fig.~\ref{FIG:Fig_ExperimFlux} the
parametrization of the flux only gives the main features of the
experimental flux. The parameters of this linear function are
determined by getting a good description of the density function,
which in this case is the ion saturation current. In
Fig.~\ref{FIG:Fig_CompTheoExp} we have plotted the ion saturation
current, Fig.~\ref{FIG:Fig_CompTheoExp}a, the averaged flow velocity
shear, Fig.~\ref{FIG:Fig_CompTheoExp}b, the ion saturation current
fluctuation, Fig.~\ref{FIG:Fig_CompTheoExp}c, and the toroidal
correlation, Fig.~\ref{FIG:Fig_CompTheoExp}d. The agreement between
the experimental data and the model description is quite satisfactory,
especially noting the extreme simplicity of the latter.

\section{Conclusions and further work}

Recently, long-distance toroidal correlations in the electrostatic
potential fluctuations have been observed in
TJ-II~\cite{Pedrosa08}. The value of the correlations increases above
the critical point of the transition for the emergence of the plasma
edge sheared flow layer. In a previous work the transition was
interpreted in terms of a simple model~\cite{CarGarPedHid06}
consisting of envelope equations for the level of fluctuations, the
mean flow shear and the averaged pressure gradient.

In the present paper we have shown that the phenomenon of
long-distance correlations requires the extension of the model so that
the effect of zonal flows is taken into account. With the addition of
an equation for the zonal flow amplitude shear the model is able to
capture the basic features of the experimental results.

The structure of the model equations is based on quasilinear
calculations of resistive pressure-gradient-driven turbulence, and we
leave for future publications the detailed calculations which should
allow to compute some parameters of the model. Herein, we have
determined those parameters by fitting the experimental data.

\begin{acknowledgments}
Part of this research was sponsored by the Direcci\'on General de
Investigaci\'on of Spain under project
ENE2006-15244-C03-01. B. A. C. thanks the financial support of
Universidad Carlos III and Banco Santander through a {\it C\'atedra de
Excelencia}.
\end{acknowledgments}


\newpage

\newpage

\newpage

\newpage

\newpage

\newpage

\newpage

\end{document}